\documentclass[aps,superscriptaddress,twocolumn]{revtex4}
\usepackage{amsmath}
\usepackage{graphicx}
\usepackage{dcolumn}
\usepackage{verbatim}
\usepackage{times}
\usepackage{subfigure}
\usepackage{bm}
\usepackage{color}
\usepackage[colorlinks,citecolor=blue,linkcolor=blue,dvipdfm]{hyperref}
\usepackage{subfigure}

\usepackage{booktabs}
\usepackage{appendix}

\begin{document}

\title{Weyl nodes and Kondo interaction in noncentrosymmetric semimetals RGaGe (R=La, Ce and Pr) with long Fermi arcs}

\author{Huan Li}
\email{lihuan@glut.edu.cn}
\affiliation{College of Physics and Electronic Information Engineering, Guilin University of Technology, Guilin 541004, China}
\affiliation{Key Laboratory of Low-dimensional Structural Physics and Application, Education Department of Guangxi Zhuang Autonomous Region, Guilin 541004, China}

\date{\today}

\begin{abstract}

In non-centrosymmetric Weyl semimetals based on rare-earth compounds, the correlation effects among $f$ electrons may play a crucial role in shaping the properties of Weyl nodes, potentially giving rise to magnetic Weyl semimetals or correlated Weyl excitations. Here, we investigates a recently identified class of magnetic rare-earth Weyl semimetals, RGaGe (R = La, Ce, and Pr). By employing a combination of density functional theory (DFT) and dynamical mean-field theory (DMFT) calculations, we demonstrate that under ambient pressure, the $f$ electrons in CeGaGe and PrGaGe are nearly fully localized. Concurrently, three inequivalent types of Weyl nodes emerge near the Fermi level due to intersections between $spd$ bands. Notably, one type of Weyl point exhibits substantial chiral separation (significantly greater than in CeAlSi and CeAlGe), leading to the formation of long and well-defined surface Fermi arcs on the (001) surface. These Fermi arcs remain well-separated by bulk states, thereby facilitating future experimental observations. In contrast, the chiral separation of Weyl points in LaGaGe is relatively modest. Upon application of volume compression, the $f$ electrons in CeGaGe progressively become itinerant and begin to obscure the Weyl nodes formed by $spd$ electrons, rendering them less accessible for direct observation. These findings suggest that in correlated materials, particularly $f$-electron systems, even when $f$ electrons do not directly contribute to the formation of topological nodes, they can still exert a profound influence on the characteristics of Weyl nodes.

\end{abstract}

\maketitle

\section{Introduction}

Since the theoretical prediction and experimental verification, Weyl semimetals have garnered significant attention due to their distinctive topological surface states and anomalous Hall effect, among other novel physical properties~\cite{Wan11}. Such materials can form in systems where either the inversion symmetry or the time-reversal symmetry is broken~\cite{Weng15,Xu15}. Among them, rare-earth compounds, owing to the presence of correlated $4f$ electrons, feature a multitude of interactions such as electron correlation, magnetism, and Kondo effect interwoven with the lattice symmetry, serving as an ideal platform for exploring exotic physical behaviors~\cite{Lai18}. For instance, in non-centrosymmetric structures like Ce$_3$Bi$_3$Pt$_4$ and CeRh$_4$Sn$_6$, the intersection of $f$ hybridized bands gives rise to a heavy-fermion type Weyl semimetal state and demonstrates novel transport properties such as nonlinear response and the anomalous Hall effect~\cite{Xu17,Cao20,Kofuji21}; while in the centrosymmetric system CeSb, the magnetism originating from $f$ electrons at low temperatures can induce the formation of Weyl points and is accompanied by a negative magnetoresistance effect~\cite{Guo17}.

In recent years, some non-centrosymmetric rare-earth compounds with tetragonal structures, RAlX (R = La, Ce, and Pr; X = Si and Ge), have been theoretically predicted to be Weyl semimetals and encompass multiple Weyl points~\cite{Puphal19,Liu21,Gao24,Wu23,Forslund25}. Particularly, these materials exhibit diverse magnetic structures at low temperatures~\cite{Yang21,Forslund25}. Consequently, as the temperature drops, the positions of Weyl points in the paramagnetic phase will shift or split under the influence of magnetism~\cite{Gao24}. Very recently, a class of structurally similar materials, RGaGe (R = La, Ce, Pr), have been experimentally discovered. X-ray diffraction experiments affirmed that they belong to the $I4_1md$ space group and exhibit ferromagnetic order at low temperatures~\cite{Patil96,Ram23,Scanlon24}. Near the magnetic phase transition temperature, these materials manifest negative magnetoresistance behavior, suggesting that they might possess Weyl semimetal characteristics~\cite{Ram23}. Nevertheless, the Weyl semimetal features of these materials still require further verification through theoretical and experimental approaches.

For Weyl semimetals formed by rare-earth compounds, the $f$ electrons can directly participate in the formation of Weyl points. For example, in Ce$_3$Bi$_3$Pt$_4$ and CeRh$_4$Sn$_6$, the contribution of $f$ electrons leads to the generation of heavy-fermion Weyl quasiparticles~\cite{Xu17,Cao20}; while in other magnetic rare-earth compounds such as CeSb and CeAlSi, the $f$ electrons in rare-earth atoms are completely localized and generate magnetic moments, and the Weyl nodes are entirely formed by the magnetic effect on the $spd$ itinerant electrons~\cite{Guo17,Sakhya23}. In fact, in the topological semimetal formed by rare-earth compounds such as CePt$_2$Si$_2$, it has been reported that the Kondo behavior of $f$ electrons can regulate the behavior of Dirac nodes, such that the Dirac quasiparticles change from light to heavy as the temperature decreases from high to low~\cite{Ma23}. Therefore, when studying the Weyl semimetal behavior of RGaGe (R = La, Ce, Pr), the behavior of $4f$ electrons must be taken into consideration to accurately describe their temperature-dependent characteristics.

Based on this backdrop, we initially computed the $f$-electron correlation behavior of rare earth atoms in CeGaGe and PrGaGe, by employing density functional theory in combination with the dynamical mean field algorithm. The results indicate that under ambient pressure, the Kondo resonance of $f$ electrons near the Fermi level is extremely weak, thus the $f$ electrons are almost completely localized and can be regarded as core electrons. Therefore, the band structure can be calculated based on $spd$ electrons via density functional theory. By integrating density functional theory with the maximum localized Wannier function fitting method~\cite{Pizzi20}, we obtained the tight-binding Hamiltonian and confirmed the existence of three types of inequivalent Weyl nodes near the Fermi level in RGaGe (R = La, Ce, Pr), and determined the specific coordinates of these Weyl points in the wave vector space. Additionally, we discovered that on the (001) surface of CeGaGe, one type of Weyl point extends a notable Fermi arc, which is well separated from the bulk states and is conducive to experimental observation. Finally, we calculated CeGaGe under volume compression using density functional theory in combination with the dynamical mean field algorithm and found that under increasing volume compression, the localized nature of $f$ electrons gradually transforms into itinerant behavior, resulting in a strong Kondo resonance peak at the Fermi level. At this juncture, there is a substantial $f$-electron spectral weight near the Weyl nodes, indicating that the possible Weyl semimetal state has transcended the realm of application of the single-electron image.

\section{DFT+DMFT calculation and $4f$ electronic correlations}

\begin{figure}[tbp]
\hspace{-0cm} \includegraphics[totalheight=1.95in]{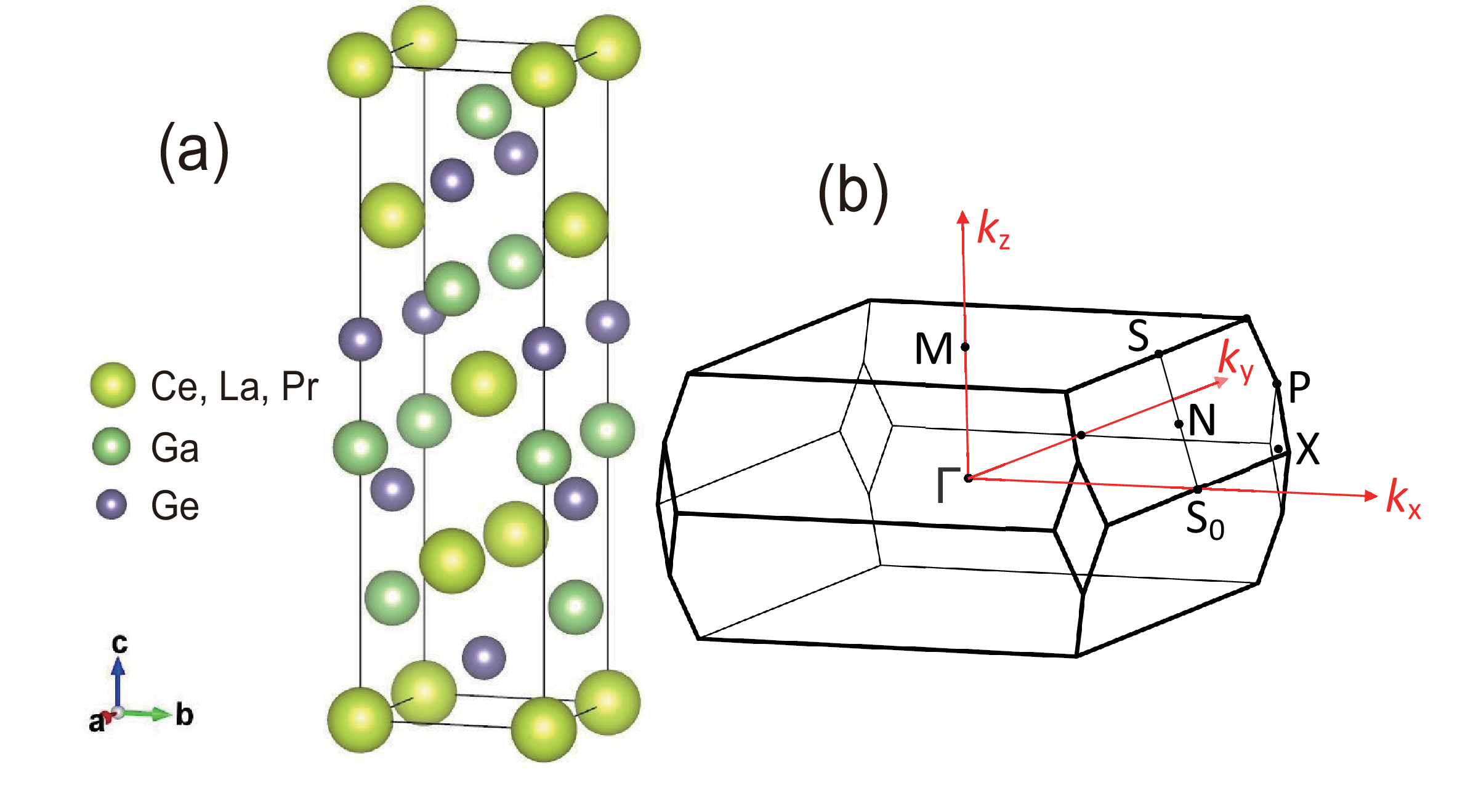}
\caption{ (a) Crystal structure of RGaGe (R=La, Ce and Pr) with $I4_1md$ space group. (b) Corresponding Brillouin zone.
}
\label{Lattice}
\end{figure}

Recently, X-ray diffraction and neutron diffraction experiments have confirmed that RGaGe (R = La, Ce, Pr) exhibits a non-centrosymmetric structure and belongs to the $I4_1md$ space group (No. 109)~\cite{Scanlon24}. Magnetic susceptibility measurements revealed magnetic transitions at approximately 6 K and 19.4 K in CeGaGe and PrGaGe~\cite{Patil96,Ram23}, respectively, while negative magnetoresistance was observed in the paramagnetic phase~\cite{Ram23}. In contrast, LaGaGe remains paramagnetic across the entire temperature range~\cite{Ram23}. Fig. \ref{Lattice}(a) illustrates the crystal structure of RGaGe, clearly showing the absence of a spatial inversion center. Fig. \ref{Lattice}(b) depicts the Brillouin zone and its high-symmetry points. Notably, RGaGe and TaAs belong to the same space group~\cite{Weng15}, sharing a consistent Brillouin zone structure, which results in similar characteristics of their Weyl nodes.

\heavyrulewidth=1bp

\begin{table}
\small
\renewcommand\arraystretch{1.3}
\caption{\label{tab1}
Correlation parameters $U$ and $J$ calculated via the doubly screened Coulomb correction approach~\cite{Liu23}, which are adopted in DFT+DMFT simulations.}
\begin{tabular*}{8cm}{@{\extracolsep{\fill}}ccccccccc}
\toprule
        &   $U$ (eV) & $J$ (eV) \\
\hline
 CeGaGe &  5.45& 0.67  \\
 PrGaGe &   6.05&  0.73 \\
 \bottomrule
\end{tabular*}
\label{tab1}
\end{table}

\begin{figure}[tbp]
\hspace{-0cm} \includegraphics[totalheight=2.4in]{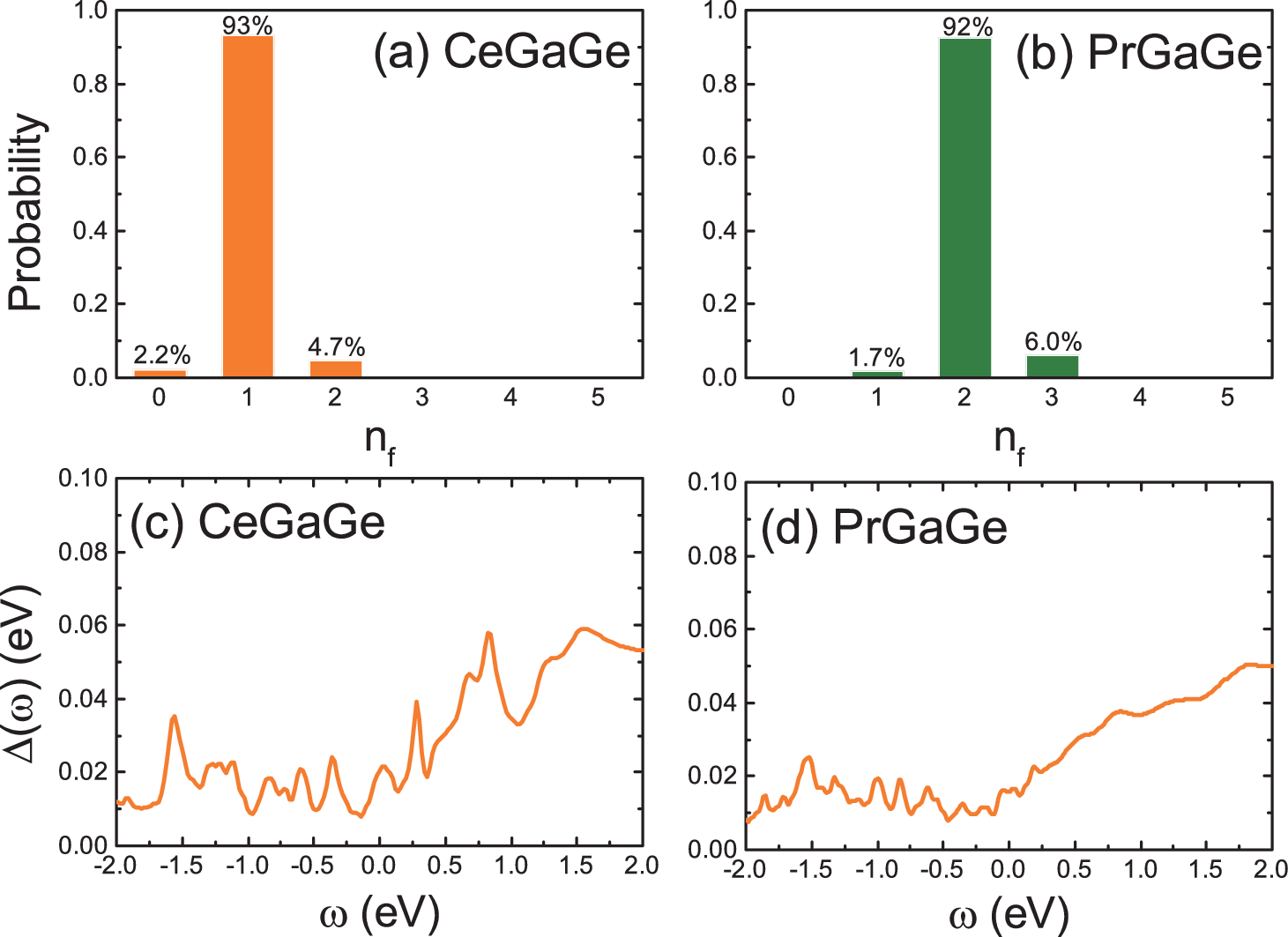}
\caption{ (a) and (b) illustrate the probabilities of Ce-4$f$ states with $0$ to $5$ electrons, in CeGaGe and PrGaGe, respectively. (c) and (d) show the impurity hybridization functions in DFT+DMFT simulations of CeGaGe and PrGaGe, respectively.
}
\label{Delta_probability}
\end{figure}

\begin{figure*}[tbp]
\hspace{0cm} \includegraphics[totalheight=4.2in]{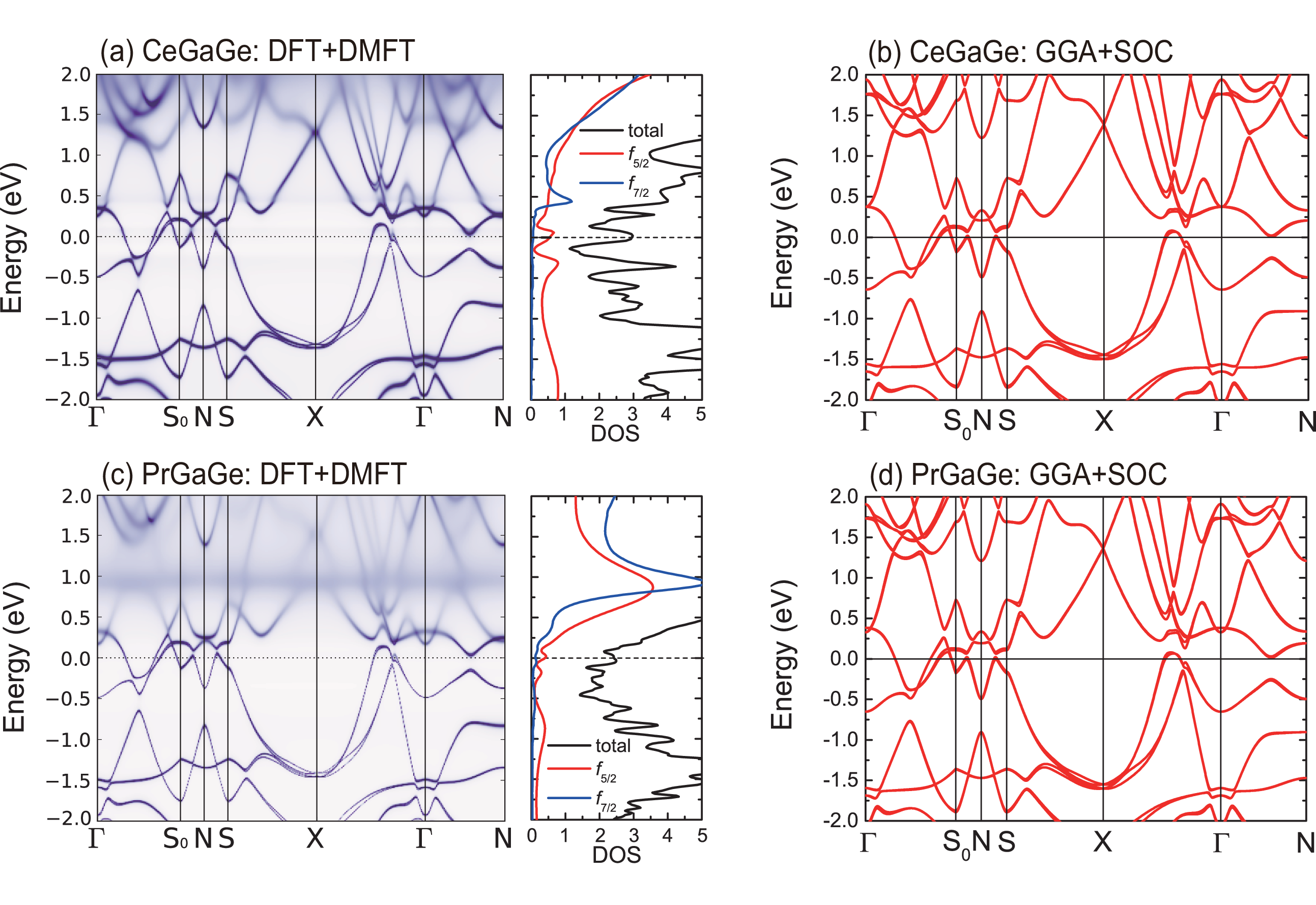}
\caption{ (a) and (c) represent the DFT+DMFT momentum-resolved spectral functions of CeGaGe and PrGaGe at 116 Kelvin, respectively. In the right patterns in (a) and (c), the total DOS, partial DOS of 4$f_{5/2}$ and 4$f_{7/2}$ states are shown by dark, red and blue lines, respectively. (b) and (d) represent the 4$f$ open-core DFT energy bands of CeGaGe and PrGaGe, respectively.
}
\label{specfunc}
\end{figure*}

Specific heat measurements indicate that the Sommerfeld coefficient of CeGaGe is significantly higher than that of CeAlGe~\cite{Scanlon24}, suggesting stronger electron correlation effects. To investigate the correlation effects of $4f$ electrons and their influence on the electronic structure in CeGaGe and PrGaGe, we employed the DFT+DMFT algorithm, combining density functional theory (DFT) with dynamical mean-field theory (DMFT), to calculate the paramagnetic phase at low temperatures. Specifically, we utilized the EDMFT package to perform full-charge self-consistent calculations for DFT+DMFT~\cite{Haule10}, incorporating spin-orbit coupling (SOC) effects, with the quantum impurity solver employing the continuous-time quantum Monte Carlo algorithm (CTQMC). For rare-earth compounds, the correlation parameters of $4f$ electrons in rare-earth elements play a critical role in determining their Kondo effect and other correlation effects. Here we calculated the on-site Coulomb repulsion energy $U$ and Hund's coupling strength $J$ of $4f$ electrons using the doubly screened Coulomb correction approach (DSCC) within Vienna \textit{ab} \textit{initio} Simulation
Package (VASP)-based DFT calculations~\cite{Liu23}. The results are presented in Tab. \ref{tab1}, showing that these parameters are comparable to those of other typical Ce-based and Pr-based compounds~\cite{Shim07,McMahan05,Galler22}.

Fig. \ref{Delta_probability} presents the DFT+DMFT results for CeGaGe and PrGaGe. Fig. \ref{Delta_probability}(a) and (b) reveal that the most probable configurations of 4$f$ electrons in Ce and Pr ions are
$4f^1$ and $4f^2$, respectively, with occupation probabilities exceeding 90\%, while other occupation numbers exhibit relatively low probabilities. The average occupation numbers of $4f$ electrons in Ce and Pr are 1.03 and 2.05, respectively, indicating that their ionic valence states closely resemble Ce$^{3+}$ and Pr$^{3+}$, with minor valence fluctuations. This suggests that the pseudopotentials of Ce$^{3+}$ and Pr$^{3+}$ can be equivalently used to describe the correct band structures via 4$f$ open-core DFT, as shown in Fig. \ref{specfunc}(b) and (d). The DFT+DMFT spectral functions (Fig. \ref{specfunc}(a) and (c)) closely match the 4$f$ open-core DFT bands over a wide energy range, demonstrating that the $4f$ electrons in Ce and Pr are nearly completely localized and do not directly participate in band formation. From Fig. \ref{Delta_probability}(c) and (d), it is evident that the impurity hybridization functions of $f$ electrons are rather weak near the Fermi level, leading to low Kondo resonance peaks. The partial density of states (PDOS) in the right panel of Fig. \ref{specfunc}(a) shows that near the Fermi level, the $f_{5/2}$ and $f_{7/2}$ peaks of Ce are separated, forming a three-peak structure similar to $\gamma$-Ce metal~\cite{Zhu20}. These peaks are quite low, positioned at the Fermi level and $\pm E_{\mathrm{soc}}$, with the splitting size determined by the spin-orbit coupling strength $\pm E_{\mathrm{soc}}$, approximately 0.4 eV. The Kondo resonance peak at the Fermi level is primarily contributed by the $f_{5/2}$ state. The PDOS of PrGaGe (right panel in Fig. \ref{specfunc}(c)) exhibits similar features but with smaller peak heights and separations compared to CeGaGe, making its three-peak structure less pronounced.

Due to the limited contribution of $f$ electrons near the Fermi level, the $f$-electron spectral weights in the spectral functions of CeGaGe and PrGaGe (Fig. \ref{specfunc}(a) and (c)) are very small, resulting in weak $f_{5/2}$ and $f_{7/2}$ hybridization bands near the Fermi level, which are almost invisible in the figure. In contrast, the upper Hubbard band of CeGaGe is broader and exhibits strong spectral weight (centered at 2.5 to 3 eV, close to $U/2$), while the lower Hubbard band is weaker (located near $-2$ eV). For PrGaGe, a strong lower Hubbard band appears near $-4$ eV for the $f_{5/2}$ state, and the upper Hubbard band near $1$ eV is contributed by both $f_{5/2}$ and $f_{7/2}$ states, resembling the behavior of Pr metal~\cite{McMahan05}. Additionally, since the atomic numbers of Ce and Pr differ by $1$, and their ionic configurations are 4$f^1$ and 4$f^2$, respectively, contributing the same number of conduction electrons. Consequently, the 4$f$ open-core DFT results for CeGaGe and PrGaGe are highly similar (see Fig. \ref{specfunc}(b) and (d)), leading to very close Weyl point coordinates. A detailed analysis is provided below.

\section{band structure and Weyl nodes}

\begin{figure*}[tbp]
\hspace{0cm} \includegraphics[totalheight=3.2in]{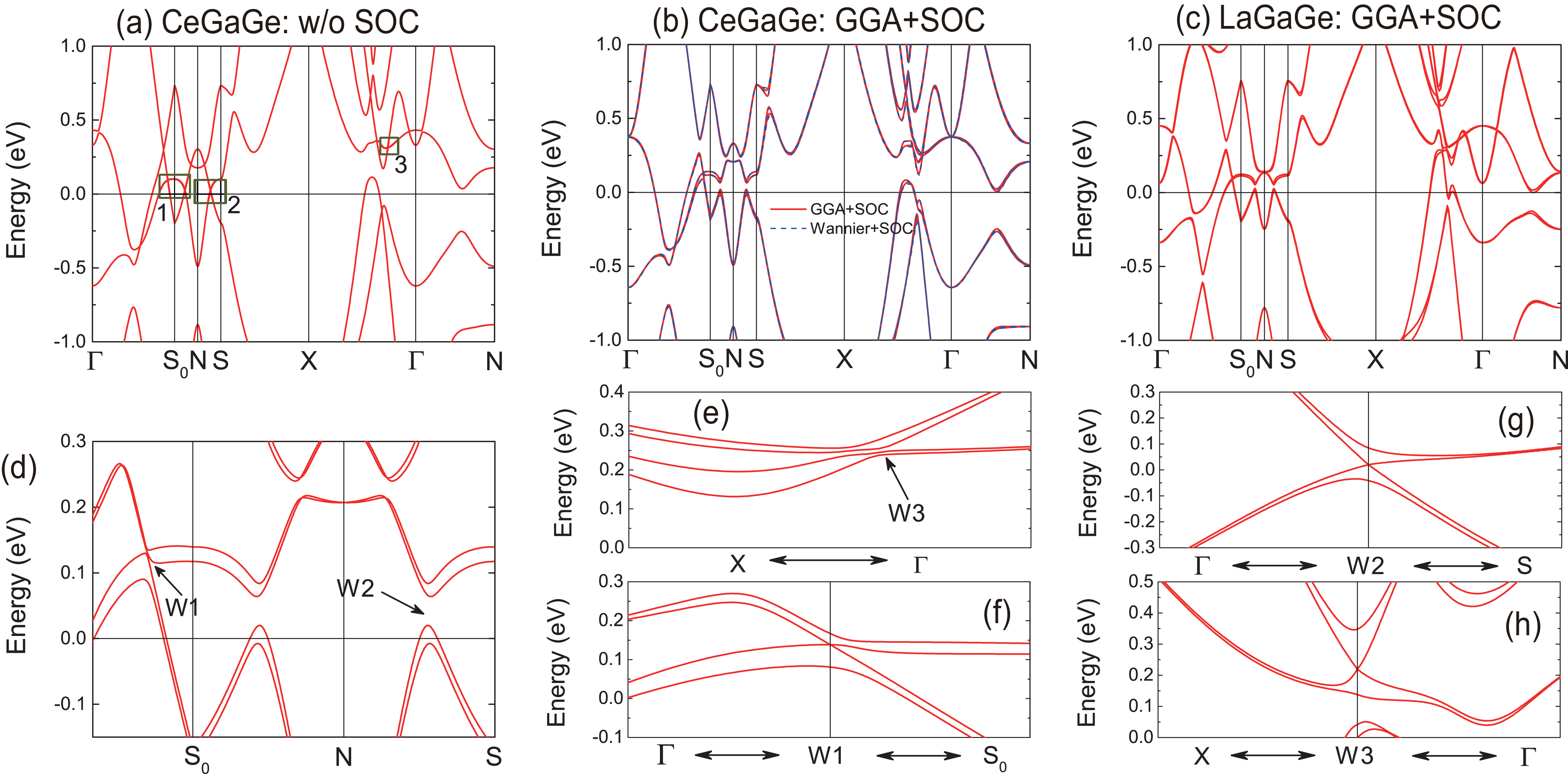}
\caption{ (a) and (b) illustrate the electronic bands of CeGaGe without and with SOC, respectively. In (b) the fitted tight-binding bands through Wannier90 interface are shown by dashed blue lines, showing good match with DFT bands. (c) GGA+SOC bands of LaGaGe. (d) and (e) are the enlarged images of GGA+SOC bands of CeGaGe along the $\mathbf{k}$ pathes in the vicinity of Weyl points. (f) to (h) plot the dispersions through the Weyl points.
}
\label{bands}
\end{figure*}

\begin{figure}[tbp]
\hspace{0cm} \includegraphics[totalheight=2.4in]{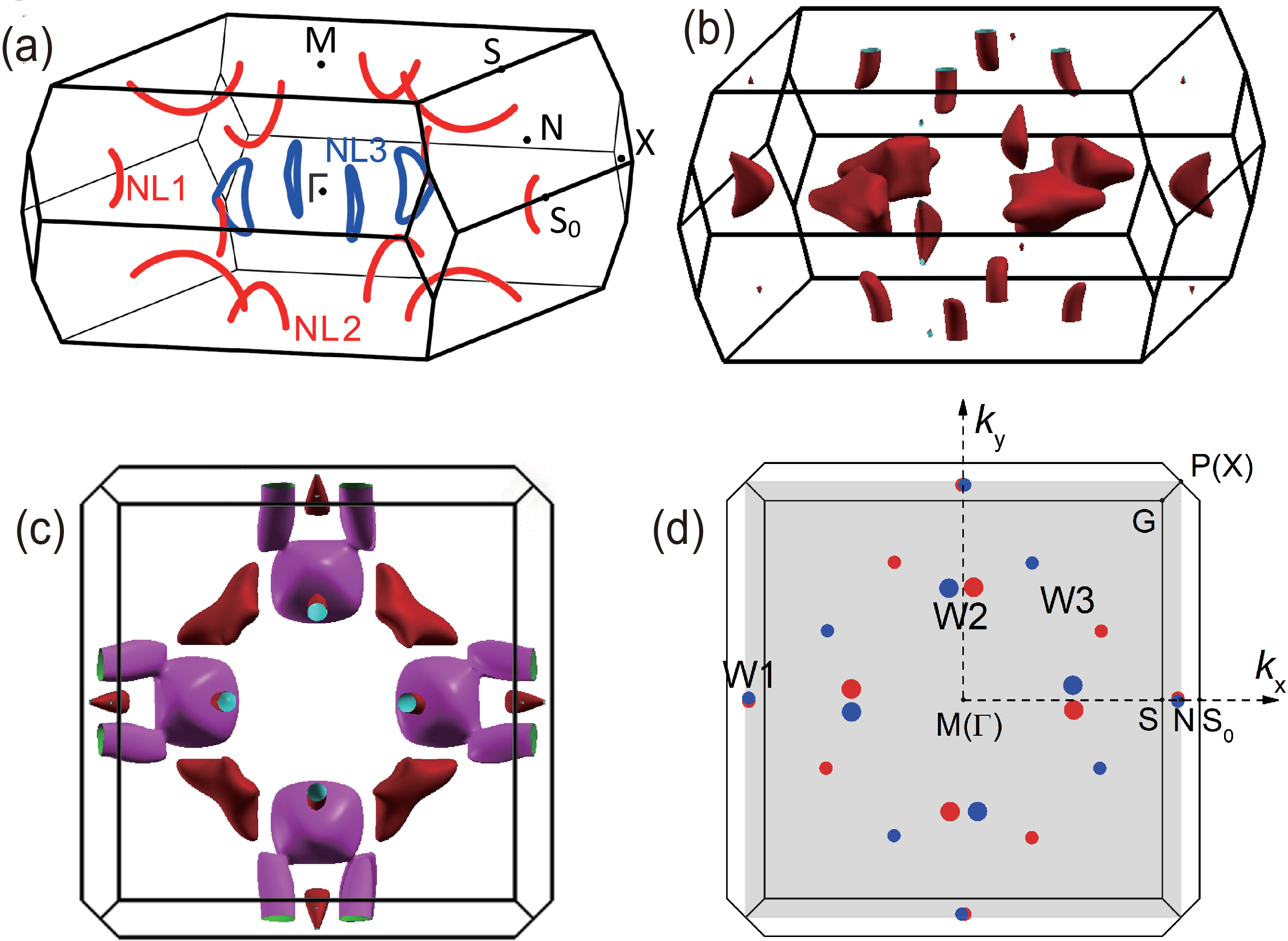}
\caption{ (a) Node lines of CeGaGe without SOC, which are indicated by the green rectangles in Fig. \ref{bands}(a). (b) Hole pockets of CeGeGe at present of SOC. (c) Bulk Fermi surfaces of CeGeGe at present of SOC, in which the electron and hole pockets are indicated by purple and red colors, respectively. (d) Top view from [001] direction of the Weyl points in CeGaGe, in which the $+1$ and $-1$ chiralities are indicated by red and blue dots, respectively, while the larger dots represent the overlap of two Weyl points with identical chirality during projection.
}
\label{BulkFS}
\end{figure}

The preceding section has demonstrated that, under ambient pressure, the $f$ electrons in CeGaGe and PrGaGe are nearly fully localized. Consequently, the $4f$ open-core DFT method is suitable for describing their band structures. Prior to calculating the Weyl nodes of CeGaGe, it is advantageous to first analyze the band characteristics without spin-orbit coupling, as depicted in Fig. \ref{bands}(a). It is evident that there are three distinct band crossings near the Fermi level (indicated by boxes in the figure), which form three closed nodal lines. The shapes of these nodal lines within the Brillouin zone are presented in Fig. \ref{BulkFS}(a), labeled as NL$_1$, NL$_2$, and NL$_3$. Specifically, NL$_1$ and NL$_2$ are located on the $k_x=0$ and $k_y=0$ planes, while NL$_3$ appears on the $k_x\pm k_y=0$ planes. The band structure incorporating spin-orbit coupling is illustrated in Fig. \ref{bands}(b), with Figs. \ref{bands}(d) and \ref{bands}(e) providing magnified views of the bands around the original nodal lines. It can be observed that small gaps have opened up at each nodal line, although band crossings persist only at specific Weyl node positions. Fig. \ref{BulkFS}(b) displays the hole-type Fermi pockets (red regions) that emerge upon inclusion of spin-orbit coupling, which show good correspondence with the nodal lines in the absence of spin-orbit coupling. Furthermore, substantial electron pockets are evident along the $k_x$ and $k_y$ axes, as highlighted in the purple regions of Fig. \ref{BulkFS}(c). In contrast, while the Fermi surfaces of CeAlSi and LaAlSi exhibit hole pockets similar to those of CeGaGe, no analogous electron pockets have been detected~\cite{Yang21,Su21,Sakhya23}.

\begin{figure*}[tbp]
\hspace{0cm} \includegraphics[totalheight=3.8in]{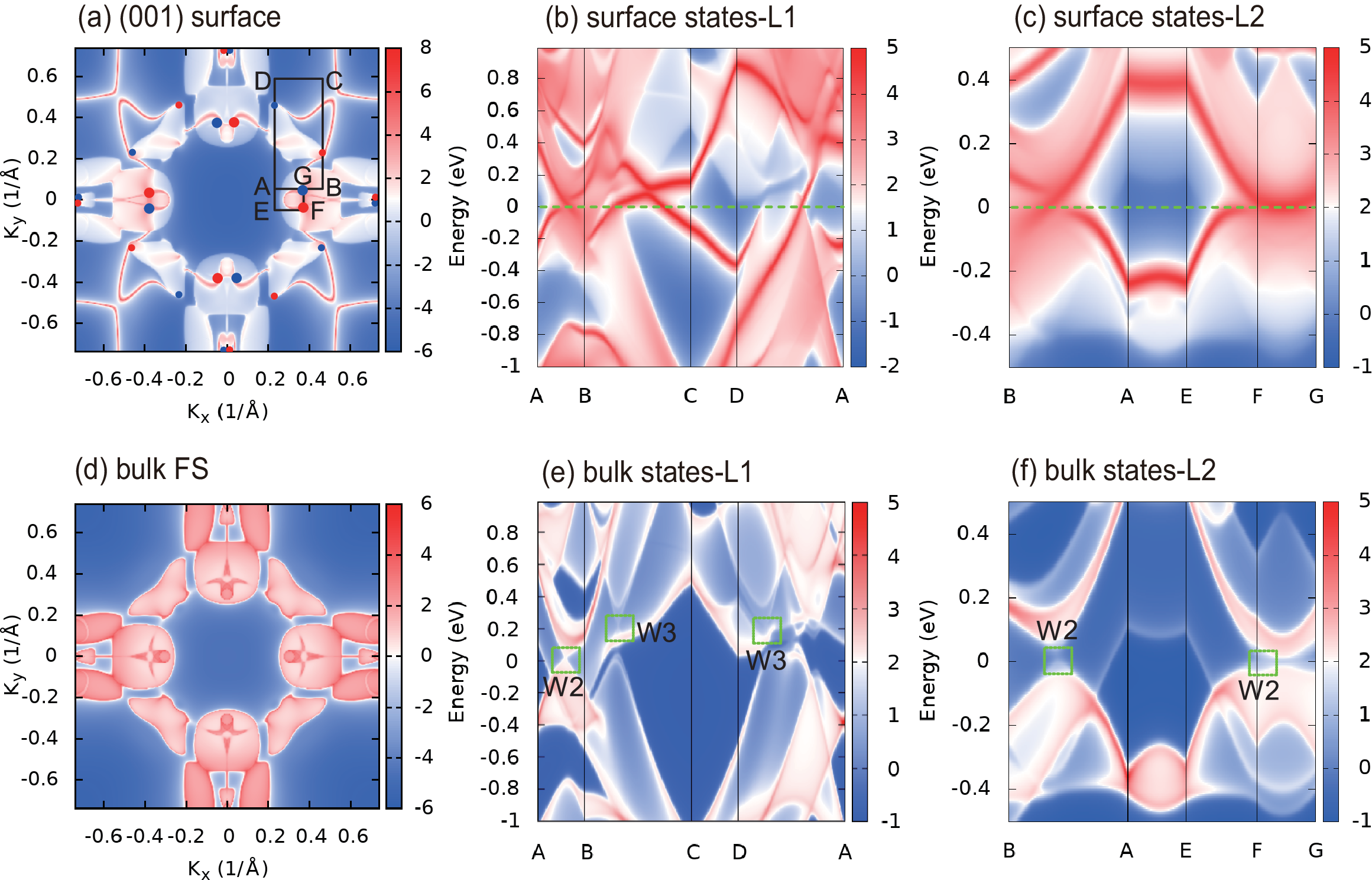}
\caption{ (a) Surface Fermi surfaces on (001) surface of CeGaGe, in which the red and blue dots represent the projective points of bulk Weyl nodes with $+1$ and $-1$ chiralities, respectively. (b) and (c) display the surface states along two closed paths showing in (a). (d) illustrates the contribution of bulk states to the Fermi surface on the (001) surface.
(e) and (f) show the surface DOS projecting to the bulk states, in which the Weyl nodes can be identified.
}
\label{Surfacearc}
\end{figure*}

To precisely determine the coordinates of the Weyl nodes, we utilized the Wannier90 program based on maximally localized Wannier functions to construct the tight-binding Hamiltonian for CeGaGe through Ce $4d$; Ga $s$, $p$; and Ge $s$, $p$ orbitals. The corresponding band structure is displayed by blue dashed lines in Fig. \ref{bands}(b), showing excellent agreement with the $4f$ open-core DFT band structure. Using this tight-binding Hamiltonian, we identified the $k$ coordinates of the Weyl nodes via the WannierTools package, with the detailed results listed in Tab. \ref{tab2}. Our analysis reveals that CeGaGe hosts three inequivalent types of Weyl nodes, and the band dispersion characteristics near these nodes are illustrated in Fig. \ref{bands}(f) to (h). It is evident that all these Weyl nodes belong to the type I. Furthermore, by calculating the Wannier charge centers of the closed surfaces surrounding these Weyl nodes, we have determined their chiralities. Fig. \ref{BulkFS}(d) depicts the distribution of these three types of Weyl nodes in the $k_x$-$k_y$ plane, where red and blue dots correspond to Weyl nodes with chiralities of $+1$ and $-1$, respectively.

Specifically, the first type of Weyl nodes (W$_1$) consists of four pairs, located on the $k_z=0$ plane and distributed along the $k_x$ and $k_y$ axes. These nodes are very close to the boundary of the surface Brillouin zone of the (001) surface (the shaded area in Fig. \ref{BulkFS}(d). The internal coordinate separation of these paired W$_1$ nodes is extremely small, with $\Delta k_x$ or $\Delta k_y$ approximately 0.003 ${\AA}^{-1}$ (see Tab. \ref{tab2}). The second type of Weyl nodes (W$_2$) appears on the $k_z=\pm0.275$ planes. The four pairs of W$_2$ nodes on each plane project onto the $k_x$-$k_y$ plane with identical coordinates and chirality, causing them to overlap. These overlapping W$_2$ nodes are represented by larger dots in Fig. \ref{BulkFS}(d). The internal coordinate separation of the W$_2$ node pairs is significantly larger than that of the W$_1$ nodes, with $\Delta k_x$ or $\Delta k_y$ approximately 0.047 ${\AA}^{-1}$ (see Tab. \ref{tab2}). The third type of Weyl nodes (W$_3$) is also located on the $k_z=0$ plane, consisting of four pairs symmetrically distributed along $k_x\pm k_y=0$. In summary, CeGaGe contains a total of 16 pairs of Weyl nodes. Among them, the spatial distribution characteristics of W$_1$ and W$_2$ nodes are similar to those of TaAs~\cite{Xu15,Weng15}, which shares the same space group and non-centrosymmetric property. Additionally, compared with CeAlSi and LaAlSi~\cite{Yang21,Su21,Sakhya23}, which have the same lattice structure to CeGaGe, the internal coordinate separation of the paired W$_3$ nodes in CeGaGe is significantly larger.

\heavyrulewidth=1bp

\begin{table}
\small
\renewcommand\arraystretch{1.3}
\caption{\label{tab2}
The $(k_x, k_y, k_z)$ coordinates corresponding to three inequivalent Weyl nodes, with chirality $\gamma=+1$. The coordinates of the remaining Weyl nodes can be determined based on the symmetry relations showing in Fig. \ref{BulkFS}(d).}
\begin{tabular*}{8.5cm}{@{\extracolsep{\fill}}ccccccccc}
\toprule
 &Weyl node 1       & Weyl node 2          &   Weyl node 3   \\
\hline
CeGaGe&(0.726, 0.003, 0)& (0.378, -0.047, 0.275)  & (0.462, 0.232, 0)   \\
PrGaGe&(0.738, 0.001, 0)& (0.376, -0.057, 0.273)& (0.459, 0.234, 0)\\
LaGaGe&(0.719, 0.001, 0) &(0.373, -0.048, 0.290) & (0.273, 0.217, 0)\\
 \bottomrule
\end{tabular*}
\label{tab2}
\end{table}

Tab. \ref{tab2} also provides the Weyl point coordinates of PrGaGe and LaGaGe. As discussed earlier, the band structure of PrGaGe closely resembles that of CeGaGe, leading to very similar Weyl point coordinates for both materials. In contrast, the band structure of LaGaGe differs significantly from the other two, particularly along the $X$-$\Gamma$ direction (see Fig. \ref{bands}(c)). This results in a notable difference in the W$_3$ node coordinates and a smaller separation of internal coordinates within the W$_3$ node pairs. It is worth noting that the Fermi surface of RGaGe (R = La, Ce, Pr) features large hole pockets and electron pockets, which partially obscure the W$_1$ and W$_2$ Weyl points, thereby complicating the observation of their topological surface states. Fortunately, the W$_3$ Weyl points are located at the edge of the hole pockets and remain uncovered, ensuring that their associated topological surface states and Fermi arcs are not masked by bulk states. For a more detailed analysis, please refer to the subsequent sections.

\begin{figure*}[tbp]
\hspace{0cm} \includegraphics[totalheight=2.2in]{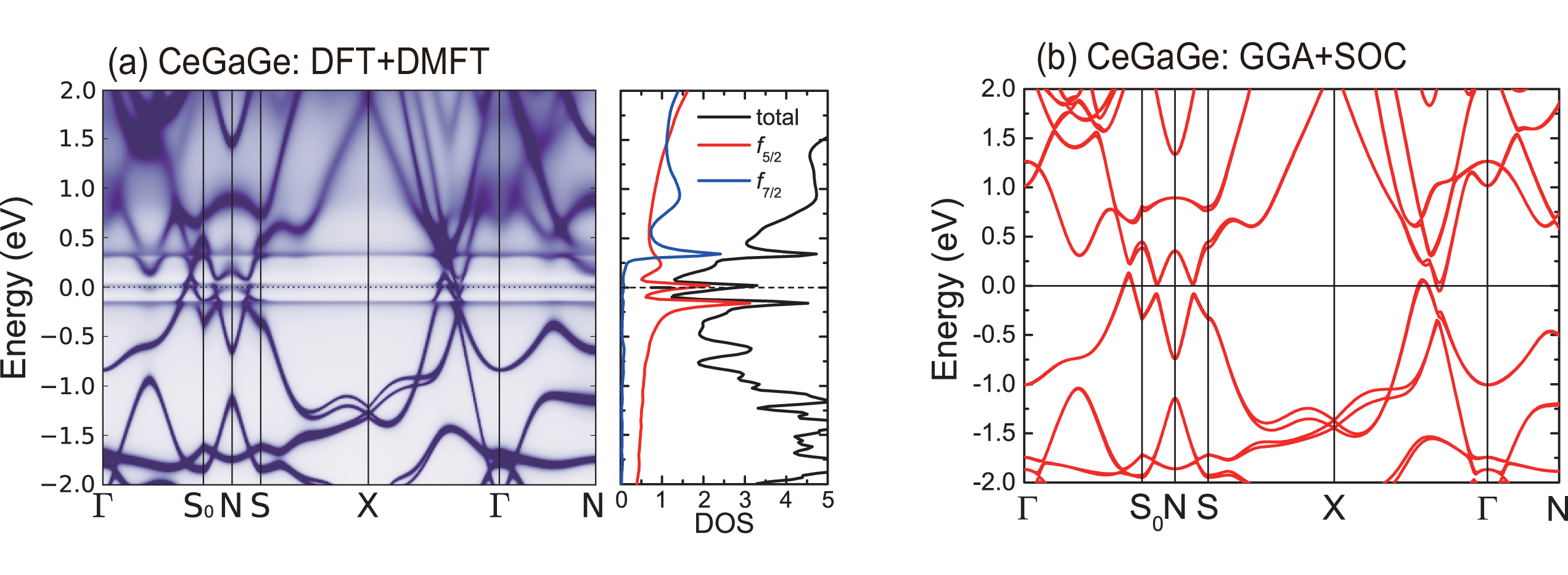}
\caption{ (a) DFT+DMFT spectral functions of CeGaGe under 20\% volume compression. In the right panel of (a), the total DOS, partial DOS of 4$f_{5/2}$ and 4$f_{7/2}$ states are shown by dark, red and blue lines, respectively. (b) represents the 4$f$ open-core DFT energy bands of CeGaGe. Under this volume compression condition, the two exhibit pronounced differences in the vicinity of the Fermi energy.
}
\label{pressure}
\end{figure*}

\section{Topological surface states and Fermi arcs}

For a particular surface of a Weyl semimetal, if the projections of a pair of Weyl points with opposite chirality do not coincide on this surface, a non-closed Fermi arc connecting this pair of Weyl points with opposite chirality will form within the corresponding surface Brillouin zone~\cite{Lv15,Xu17}. Taking the RGaGe (R = La, Ce, Pr) materials as an example, the coordinates of the Weyl points have been determined previously, and it has been found that the overlap of these Weyl points on the (001) surface (i.e., perpendicular to the $k_z$ direction) is the least. As the (001) surface is of a square structure (see Fig. \ref{Lattice}(a)), its corresponding surface Brillouin zone is also square (as shown in the shaded area in Fig. \ref{BulkFS}(d)), and the high-symmetry point $N$ of the bulk Brillouin zone (see Fig. \ref{Lattice}(b)) is precisely located at the edge of the surface Brillouin zone. The projection of the Weyl points on this surface is depicted in Fig. \ref{BulkFS}(d), where the two W$_2$ Weyl points with the same chirality overlap and are represented by larger dots, while the W$_1$ and W$_2$ Weyl points do not coincide.

Based on the tight-binding Hamiltonian obtained from the previous fitting, we utilized the WannierTools program~\cite{Wu17} to calculate the layered slab structure of the CeGaGe material with 40 layers of unit cells stacked perpendicular to the $z$ direction with open surfaces, in order to simulate the surface states and surface Fermi arcs on the (001) surface of the real crystal (see Fig. \ref{Lattice}(a)). The projection of the Fermi surface to the inner layers (i.e., the bulk state) is shown in Fig. \ref{Surfacearc}(d), which is highly consistent with the projection of the bulk Fermi surface along the $k_z$ direction (Fig. \ref{BulkFS}(c)), indicating that it is reasonable to employ the slab model for calculating the surface states. From the surface Fermi surface pattern of the slab structure (Fig. \ref{Surfacearc}(a)), it can be observed that in addition to the Fermi surface contributed by the bulk state (the grayish-white area in the figure), there are also Fermi arcs originating from the projection points of the Weyl nodes and connecting a pair of Weyl nodes with opposite chirality. Noted that the breaking of $C_4$ screw symmetry by the (001) surface leads to distinct characteristics of the surface states along $k_x$ and $k_y$ directions. Notably, the W$_3$ Weyl point is precisely located at the edge of the bulk Fermi surface, and the Fermi arc emitted therefrom is particularly long and not masked by the bulk Fermi surface, which will be conducive to experimental observation. In contrast, due to the small separation between the pair of W$_1$ Weyl points with opposite chirality, the resulting Fermi arc is very short, while the W$_2$ Weyl point is completely covered by the bulk Fermi surface, and thus its corresponding Fermi arc is interwoven with the bulk state. Additionally, the phenomenon of long Fermi arcs emitted from the W$_3$ Weyl point similar to that in CeGaGe has not been reported in similar materials such as CeAlSi and LaAlSi~\cite{Sakhya23}. Moreover, closed Fermi surfaces were observed at the corners in Fig. \ref{Surfacearc}(a), which do not intersect Weyl points, similar phenomena have also been reported in TaAs~\cite{Huang15}, which may result from the generation of certain topological invariants during the annihilation of Weyl points, leading to the transition into a topological insulator state hence the formation of these surface states.

To investigate the surface states on the (001) surface more deeply, we further calculated the surface state distribution along two different paths. Both paths pass through the projected positions of the W$_2$ or W$_3$ points (marked by the boxes in Fig. \ref{Surfacearc}(a)), and the corresponding surface states are displayed in Figs. \ref{Surfacearc}(b) and (c), while the projections to inner layers (i.e., bulk states) are shown in Figs. \ref{Surfacearc}(e) and (f). The existence of Weyl cones can be clearly discerned from the projections to bulk states, such as the W$_2$ Weyl cone along the AB direction and the W$_3$ Weyl cones along the BC and DA paths (marked by the green boxes in Figs. \ref{Surfacearc}(e) and (f)). Simultaneously, from Figs. \ref{Surfacearc}(b) and (c), it can be seen that these Weyl cones give rise to intense surface state dispersions, such as the dispersion starting from the W$_3$ point between BC and ending at another W$_3$ point between DA, and the dispersion starting from the W$_2$ point between BA and ending at the W$_2$ point at F. These surface state dispersions correspond to the Fermi arcs in Fig. \ref{Surfacearc}(a).

\section{correlation under compression}

It has been demonstrated in the preceding text that, under ambient pressure conditions, the $f$ electrons of rare earth elements in CeGaGe and PrGaGe are localized and do not directly contribute to the formation of energy bands. As a result, the Weyl points are primarily constructed by the bands formed by $spd$ conduction electrons. For rare-earth compounds, theoretical and experimental studies have shown that when external or internal conditions vary, $f$ electrons can transform between localized and itinerant states. For example, decreasing temperature enhances the Kondo resonance effect of $f$ electrons, leading to the formation of a Kondo coherent state at low temperatures~\cite{Shim07}. In contrast, volume compression can gradually transform originally localized $f$ electrons into itinerant ones, forming heavy-fermion energy bands~\cite{Lu16,Li24}. In the case of CeGaGe and PrGaGe, due to their magnetic characteristics at low temperatures, $f$ electrons remain approximately localized across the entire temperature range under ambient pressure~\cite{Patil96,Ram23,Scanlon24}. However, under applied pressure, $f$ electrons may exhibit a tendency toward itinerancy, which further leads to the formation of heavy-fermion bands thus significantly impacts the electronic structure.

To examine the effects of volume compression on the behavior of $f$ electrons and its influence on Weyl points, we calculated the DFT+DMFT spectral function and partial DOS of CeGaGe under a 20\% volume compression ratio. The results are presented in Fig. \ref{pressure}(a). It is evident that the three DOS peaks near the Fermi energy and at $\pm E_{\mathrm{soc}}$ are substantially enhanced compared to those under ambient pressure conditions (see Fig. \ref{specfunc}(a)), forming three remarkable $f$-electron flat bands in the spectral function pattern. Under the influence of these $f$-electron bands, the remaining $spd$ conduction electron bands in the spectral function differ significantly from the 4$f$ open-core DFT bands without considering $f$ electrons (Fig. \ref{pressure}(b)) near the Fermi energy. Specifically, the band crossings where Weyl points originally appeared are traversed by the $f$-electron flat bands, causing the positions previously defining Weyl points to be largely obscured by $f$-electron weight. Consequently, the Weyl points are no longer well-defined. Therefore, beyond a certain volume compression threshold, the itinerancy of $f$ electrons may disrupt the Weyl semimetal state in CeGaGe and PrGaGe based on the single-electron approximation. At this stage, whether the system retains a Weyl semimetal character, how its topological properties differ from non-interacting Weyl semimetals, and whether anomalous Hall effects, negative magnetoresistance, or other exotic transport phenomena emerge require further investigation within a theoretical framework incorporating the interacting $f$ electrons. These findings also underscore the importance of fully accounting for $f$-electron correlations when exploring topological states in rare earth compounds.

\section{conclusion}

In summary, we utilized the DFT+DMFT method, which integrates density functional theory (DFT) and dynamical mean-field theory (DMFT), to investigate the recently experimentally reported non-centrosymmetric candidate Weyl semimetal materials RGaGe (R = La, Ce, Pr). First, we calculated the correlation parameters of these materials using the double-counting correction method and incorporated them into the DFT+DMFT framework. The results indicate that under ambient pressure, the $4f$ electrons of rare earth atoms in CeGaGe and PrGaGe are nearly fully localized, allowing the $4f$ electrons to be treated as core electrons for calculating their electronic structures based on the DFT band structure. By employing maximally localized Wannier functions for fitting, we constructed the tight-binding Hamiltonians of these materials and determined their Weyl point coordinates. Our study reveals that these materials host three types of Weyl points, totaling 16 pairs of Weyl points. Furthermore, we computed the surface Fermi arcs on the (001) surfaces of these compounds and found that one type of Weyl point in CeGaGe is located at the edge of the bulk Fermi surface, emitting significantly extended Fermi arcs. Compared with structurally analogous Weyl semimetals RAlX (R = La, Ce, Pr, X = Si, Ge) reported in the literature, the bulk Fermi surfaces of RGaGe (R = La, Ce, Pr) contain additional large electron pockets. Moreover, one representative Weyl point pair exhibits a larger wave vector separation, leading to significantly longer Fermi arcs than those in RAlX. This may result in more pronounced anomalous Hall effects and other exotic quantum phenomena~\cite{Forslund25}. Additionally, we observed that under volume compression, the $4f$ electrons in CeGaGe gradually become itinerant, causing the Weyl cones contributed by $spd$ itinerant electrons to become less distinguishable. In this scenario, whether the topological properties of the Weyl semimetal are preserved and how they evolve require further theoretical investigation.

\acknowledgments
H. Li acknowledgements the supports from National Natural Science Foundation of China (No. 12364023), and Guangxi Natural Science Foundation (No. 2024GXNSFAA010273).

\end{document}